\documentclass [12pt,twoside]{article}           

\usepackage{epsfig,times,lscape}
\usepackage[usenames]{color}

    \ifnum\count102=1

    \fi

    \ifnum\count102=2
\fi

\newcommand{\nc}{\newcommand}

     \ifnum\count101=1
\nc{\qI}[1]{\section{{#1}}}
\nc{\qA}[1]{\subsection{{#1}}}
\nc{\qun}[1]{\subsubsection{{#1}}}
\nc{\qa}[1]{\paragraph{{#1}}}

\def\qpar{\vskip 2mm plus 0.2mm minus 0.2mm}
\def\qL{\hfill \break}
     \fi 

      \ifnum\count101=2
 \nc{\qI}[1]{\parindent=0mm \vskip 8mm 
{\centerline{\LARGE \color{red}#1}}\vskip 3mm}
\nc{\qA}[1]{\vskip 2.5mm \noindent {{\bf        #1}} \vskip 1mm
\parindent=0mm}
 \nc{\qun}[1]{\vskip 1mm \noindent {\sl #1 }\quad }

\def\qL{\hfill \break}
\def\qpar{\vskip 2mm plus 0.2mm minus 0.2mm}

      \fi

\def\qth{\vrule height 12pt depth 0pt width 0pt}
\def\qtb{\vrule height 0pt depth 5pt width 0pt}

\nc{\qfoot}[1]{\footnote{{#1}}}

      \ifnum\count101=1
\def\qbu{\hfill \par \hskip 6mm $ \bullet $ \hskip 2mm}
\def\qee#1{\hfill \par \hskip 6mm (#1) \hskip 2 mm}
      \fi
      \ifnum\count101=2
\def\qbu{\hfill \par \hskip 4mm $ \bullet $ \hskip 2mm}
\def\qee#1{\hfill \par \hskip 4mm (#1) \hskip 2 mm}
      \fi

\def\qparr{ \vskip 1.0mm plus 0.2mm minus 0.2mm \hangindent=10mm
\hangafter=1}

     \ifnum\count101=1 
 
     \fi
     \ifnum\count101=2

  \def\qcitb#1{\noindent \hbox to 102mm{\hfill \small #1} \vskip 1mm}
      \fi

 \def\qpages#1{\count102=0{\loop\advance\count102 by 1
 \null \vfill\eject \ifnum\count102<#1 \repeat}}

\def\qn#1{\eqno \hbox{(#1)}}

\def\qth{\vrule height 12pt depth 0pt width 0pt}
\def\qtb{\vrule height 0pt depth 5pt width 0pt}

\def\qv{\vskip 0.1mm plus 0.05mm minus 0.05mm}

\def\qhw{\hskip 1.5mm}
\def\qleg#1#2#3{\noindent {\bf \small #1\qhw}{\small #2\qhw}{\it \small #3}\qv }
\def\qvec#1{\overrightarrow{#1}}

\begin{document}
\thispagestyle{empty}



\markboth{{\sl \hfill  \hfill \protect\phantom{3}}}
        {{\protect\phantom{3}\sl \hfill  \hfill}}

\color{yellow} 
\hrule height 40mm depth 1mm width 170mm 
\color{black}
\vskip -33mm
\centerline{\bf \LARGE \color{blue} How can one detect}
\vskip 5mm
\centerline{\bf \LARGE \color{blue} the rotation of the Earth
``around the Moon''? }
\vskip 5mm
\centerline{\bf \LARGE \color{blue} Part 3: With a simple gravity pendulum}
\vskip 16mm
\centerline{\bf \large 
Marcel B\'etrisey$ ^1 $, Bertrand M. Roehner$ ^2 $ }

\large
\vskip 8mm

{\bf Abstract}\quad 
The attraction of the Moon on objects at the surface of the Earth
gives rise to a so-called tidal force which is of the order of
$ 1/10,000,000 $ times the gravitational force of the Earth.
For instance, when the Moon is located between the Earth and the Sun
(new moon) the distance from a given terrestrial location 
to the Moon is shorter
at noon than at midnight. This reduces the gravitational acceleration
and therefore increases the period of a simple pendulum by a small
amount. Although the change is of the order of $ 0.1\mu $s
it appears that it can be detected. We give some preliminary results
and discuss how the accuracy can be further improved.
It is hoped that the present paper
will encourage new experiments in this direction.
\vskip 6mm

\centerline{\it 1 February 2012.
Preliminary version, comments are welcome}
\vskip 6mm

{\normalsize Key-words: Moon-Earth, 
tidal force, gravity, pendulum, period}
\vskip 15mm

{\normalsize 
1: ``MARCEL B\'ETRISEY CR\'EATIONS'', Sion, Valais, Switzerland.\qL
2: Department of Systems Science, Beijing Normal University, Beijing,
China. 
On leave of absence from the ``Institute for Theoretical and
High Energy Physics'' of University Pierre and Marie Curie, Paris,
France. \qL
Email address: roehner@lpthe.jussieu.fr.
}

\vfill \eject

\qI{Introduction}

The methods presented in the first two parts of this 
series of papers were both based
on detecting the {\it angular velocity vector} of the rotation 
under consideration. This method of 
detection is independent of the size, attraction or
nature of the celestial body which provokes the rotation. 
The
only variable which matters is the magnitude of the angular velocity.
In this framework
the movement around the Moon is 13 times easier
to detect than the rotation around the Sun in spite of the fact that 
the mass of the Sun is $ 2.7\ 10^7 $ times larger.
\qpar

There is another class of detection methods which rather focus
on gravitational attraction.
Obviously any movement that takes place on the Earth is
affected by the attraction of other celestial bodies
(of which the Moon and the Sun are the most important)
but this influence may be more or less easy to detect.
\qpar

The simplest way
to detect the gravitational influence of the Moon and Sun
is to observe the sea tides. 
Unfortunately,
tides have a fairly complicated connection with the gravitational
attraction because they are also dependent on the size 
and depth of the lakes, seas or oceans under consideration.
Offshore, in the deep ocean, the difference in tides is of the order
of 50\ cm to one meter,
but in bays and estuaries it can reach several meters.
Is it possible to find a physical
system for which the connection would be simpler to analyze?
\qpar

In 1913-1914 Albert A. Michelson set up an experiment which
allowed him to observe tidal waves in the laboratory (Michelson 1914).
A pipe, half-filled with water, 
166 meter long and 15 centimeter in diameter  
was laid in a 2-meter deep trench cut in a East-West direction.
At both end the pipe was sealed  with a glass wall.
The level of the water was read with the help of a
micrometer microscope. Level changes of about 20 micrometer were
observed. Because the pipe was 2 meter deep under ground
dilatation due to temperature changes was negligible. The
level changes were observed over a period of several months
and were well in agreement which what was expected%
\qfoot{Except for the fact that the magnitude of the changes was only
70\% of what would have been observed if the Earth had been perfectly
rigid. As a matter of fact, the main purpose of the
experiment was to estimate the rigidity of the Earth.}%
.

\qpar
In the mid-19th century
direct observation of tidal forces was attempted with
the help of an ``horizontal'' pendulum. Such a pendulum is
similar to a door whose axis of rotation would not be exactly
vertical and would be able to rotate with very small friction.
Such a device is very sensitive to small changes in the direction
of the axis of rotation which is why it is used 
up to the present day to set up seismometers.
Because the arm of such a pendulum is subject to a tiny gravity
$ g\sin \epsilon $ (where $ \epsilon $ is the angle of the axis
of rotation with the vertical) it is sensitive to very small forces
such as tidal forces. Needless to say, such a device is very sensitive
to external vibrations or thermal dilatation effects.
\qpar

An obvious candidate for the detection of small gravity changes
induced by tidal forces
is the simple gravity pendulum. 
In contrast to the Foucault pendulum which has two rotational
degrees of freedom, the simple pendulum moves around a fixed
axis of rotation and has therefore only one degree of freedom.
Back in the 18th and 19th 
century the simple pendulum has played an essential role in the study
of many physical effects%
\qfoot{In this connection it can be recalled that the definition
of the meter was chosen to be
almost equal to the length $ L $ of the so-called
``seconds pendulum'' which is  a pendulum whose 
period is precisely two seconds, one second for a swing in 
one direction and one second for the return swing: $ L=0.994 $ m.
As a matter of fact, before being defined as
$ 10^{-7} $ times the distance from the Equator to the North Pole,
the meter had been defined as the length of a seconds pendulum.}%
.
Can it be used to detect the effect of the gravitational field
of the Moon? 
\qpar

At the center of the Earth
the attraction of the Moon $ F_M(r) $ is exactly balanced 
by the centrifugal force due to the rotation of the Earth ``around the
Moon'' (in fact around the center of mass of the Earth-Moon system)
but at points located at the surface of the Earth
the two forces do not cancel one another exactly.
The difference is called the tidal force due to the Moon%
\qfoot{There is a similar tidal force due to the Sun.}%
. 
The order of magnitude of the corresponding
acceleration is:
$$ a_t=2GMr/R^3 $$ 
where:\qL
$ G $: constant of universal attraction, $ G=6.6\ 10^{-11} $ (SI
units)\qL
$ M $: mass of the Moon, $ M=7.3\ 10^{22} $ kg \qL
$ r $: radius of the Earth, $ r=6,400 $ km \qL
$ R $: distance from the Moon to the Earth, $ R=384,000 $ km 
\qpar

Numerical computation gives: $ a_t= 1.1\times  10^{-7}\ g $.
This small difference in gravity  will change the
period $ T_0=2\pi \sqrt{L/g} $ of a pendulum by the following amount:
$$ T=2\pi\sqrt{{ L\over g+a_t }}=T_0\left(1-0.5{ a_t\over g }\right) $$

As a result, the period of a seconds pendulum for which $ T_0=2 $ s
will be changed by about $ 0.2\mu $s. The question is 
whether such a small difference can be measured. 
\qpar
We will see that the
answer is ``yes''. 
In the following sections we give some preliminary results and
explain what can be done in order to improve the accuracy
of the measurement.
\qpar

We considered the ``new moon'' situation for the sake of
simplicity but it is clear that a similar experiment can be done
also in other configurations. For instance once the Moon has achieved
one fourth of a revolution it will be in the so-called ``first
quarter'' position. In this situation the distance between the Moon
and the surface of the Earth will be shortest for the locations where
the (solar) time is 6pm and it will be farthest for the locations
where the time is 6am.  

%
\begin{figure}[htb]
 \centerline{\psfig{width=10cm,figure=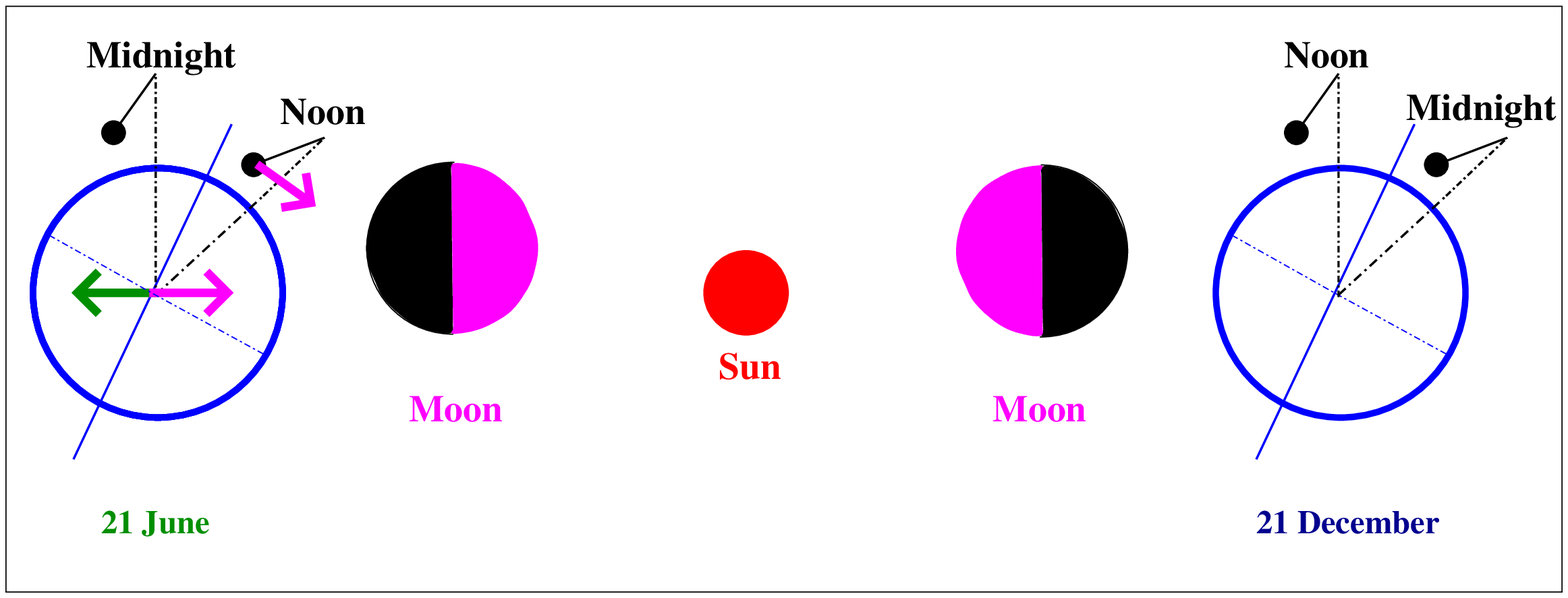}}
\vskip 3mm
\qleg{Fig. 1: Tidal force on a pendulum at noon and midnight.}
{At the center of the Earth the attraction of the Moon (in magenta)
is exactly balanced by the centrifugal force (in green) due to
the rotation of the Earth in the Earth-Moon
system. At the surface of the Earth the cancellation is no longer
an exact one however. 
The (small) difference is the so-called tidal force.
Its order of magnitude is about $ 10^{-7} $ times the Earth
gravity acceleration $ g=9.81 \hbox{m/s}^2 $.
As a result, the periods of a pendulum at noon and midnight
will differ by a small amount which is of the order of 
$ 0.2\mu $s for a pendulum whose period is 2 s. For the sake of
simplicity the diagram shows two specific 
configurations where
the Moon, the Sun and the axis of rotation of the Earth are in the
same plane. Of course, the same effect also exists 
in other configurations
although the exact amount of the difference between the longest and
shortest periods may be slightly different.}
{}
\end{figure}
%

\qI{Implementation of the experiment}

The description of the tidal force (Fig. 1) suggests a
procedure for carrying out the experiment. 

\qA{Noon-midnight difference}

In a situation where
the Moon is located between the Earth and the Sun (which is called
``new moon'' or more appropriately ``dark moon'') a person located on
the Earth will be closest to the Moon at noon and farthest at
midnight. As the Moon's attraction reduces the gravity,
$ g $ will be smaller at noon
and the period of a pendulum should therefore be larger.
The question is how to organize the measurements
so that the best accuracy can be achieved.

\qA{How many periods should one measure?}

An important uncertainty comes from the time measuring process.
A quartz clock may have a short-term
accuracy of about $ 1 \mu $s. However 
the device which starts and stops
the clock (usually an infrared detector)
may be less accurate. In order to limit the number 
of starts and stops
it is best to measure the time
for a large number of swings. 
Thus, if the interval during which the
measurement is performed covers one hour at noon and one hour
at midnight, one may in each case
perform for instance 6 measurements, each of which
will include 600 swings of 1s (i.e. 300 periods of 2s).
The $ 0.1\mu $s accuracy needed on one swing will translate
into a $ 60\mu $s requirement for the series of 600 swings.
The 6 successive measurements will allow to check the
stability of the pendulum's swings
and the reliability of the measurement process.

\qA{The damping issue}
At this level of precision, the amplitude plays an important role
even for small angles. In the small angle approximation the
relationship between the period and
the amplitude $ \theta_1 $ is the following (Cabannes 1966 p. 203):
$$ { T-T_0\over T_0 } \sim { \theta_1^2\over 16 } \qn{D1} $$

For an angle of 1 degree and $ T_0=2 $ s one gets $ T-T_0 = 38\mu $s.
In words, this means that for an amplitude of 
one degree the period is 38 microseconds longer than the ideal period
$ 2\pi\sqrt{L/g} $ of a simple pendulum.
In order to make the pendulum move during one hour without
too much damping taking place there are two possible options.
\qbu The first option is to reduce the friction as much as
possible.
\qbu The second option is to maintain the amplitude of the
movement through a propulsion device which gives back to the
pendulum the energy that it loses through friction.
\qpar

Two pendulums designed by Mr. Marcel B\'etrisey%
\qfoot{On http://www.betrisey.ch/econti.html this pendulum
can be found in the section ``Radiometric clocks''; the names
of these specific clocks are ``Chronolithe'' and ``Conti''.}
meet both requirements. 
They have a mass of 4 kg and a period of 2 s and are
contained in a glass tube in which a low air pressure of 1000 Pa
is maintained.
An initial
swing with an amplitude of 2 cm (which corresponds to an
angle of 1.15 degree) can keep them in movement for 
several hours: 6 hours for ``Chronolithe'' and 24 hours for ``Conti''.
Moreover, the amplitude can be maintained by using a ``light engine''
based on the same principle as Crookes's radiometers.

\qI{Preliminary results}

In this section we give the results already obtained.
They should be considered as preliminary 
and open to improvement for the following reasons:
\begin{table}[htb]

\centerline{\bf Table 1\quad Half period -1 of a seconds pendulum at noon
and midnight (10-13 April 2002)}
{\small
\vskip 3mm
\hrule
\vskip 0.5mm
\hrule
\vskip 2mm
$$ \matrix{
 \hbox{Trial} \hfill & \hbox{Noon} & \hbox{Noon} & \hbox{Midnight} &
\hbox{Midnight} & \hbox{\bf Noon-midnight} & \hbox{\bf Noon-midnight}\cr
  & m & \sigma & m & \sigma & m & \sigma \cr
\qtb
  & [\mu s] &[\mu s]  & [\mu s] & [\mu s]&  [\mu s]&  [\mu s]\cr
\noalign{\hrule}
\qth
1& -1.49 & 11.1 & -1.57 & 12.0 & 0.07 & 1.22\cr
2& -3.71 & 17.1 & -2.88 & \phantom{0}9.9 & -0.82& 1.42\cr
3& -0.25  & 12.4 & -3.85 & \phantom{0}9.2 & \phantom{-}3.61& 1.14 \cr
4& -0.12 & 16.9 & -2.05 & 10.9 & \phantom{-}1.94 & 1.47\cr
\qtb 
\hbox{\bf Average} & & 14 & & 10 & 1.2\pm 0.7& \cr
\noalign{\hrule}
} $$
\vskip 1.5mm
Notes: Each swing was detected by an infrared sensor and the computer
printed the average half-period every 10 swings that is to say
about every 10 seconds.
The two time intervals 
were one-hour intervals centered at noon and midnight,
that is to say 11:30-12:30 and 23:30-00:30. In each of these intervals
there were 360 measurements: $ m $ denotes the average of these
measurements and $ \sigma $ their standard deviation. \qL
As this
experiment was done in time of new moon (new moon was on 12 April
2002) one expects $ g $ to be smaller at noon which means that
the period should be longer than at midnight. The expected
difference for the half-period is $ T'_{12}-T'_{24}=0.09\mu $s
which decomposes into $ 0.066\mu $s for the Moon and $ 0.028\mu $s
for the Sun. \qL
The overall average and standard deviation can be computed by
line or by column. For the average the two computations give of course
the same result. For the standard deviation one gets two 
slightly different estimates, namely
$ 1.97/\sqrt{4}=0.98\mu $s for the $ \sigma $ of
the averages and $ 0.58\mu $s for the average of the
  $ \sigma $.\qL
Because
the error-bar which was used here is $ \pm \sigma $. 
it means that the probabilistic level of confidence is 0.68
and not 0.95 as would be the case for $ \pm 2\sigma $.\qL
{\it Source: http://www.betrisey.ch/eindex.htm.}
\vskip 2mm
\hrule
\vskip 0.5mm
\hrule
}

\end{table}
%
\qbu The trigger device which starts and stops the clock was
an infrared device which is known to be less accurate as the
clock itself.
\qbu Instead of measuring the time for 600 swings 
(as suggested above) it was measured for subsets of only 10 swings.
This required the trigger device to work much more often
and therefore amplified the inaccuracy due to this device.
\qpar

The results obtained for 5 noon-midnight comparisons are
summarized in Table 1.
Although the overall average (last line of the table) shows,
that, as expected, 
the period is longer at noon than at midnight the error bar is
still too large to make this result really conclusive.
As suggested in the discussion, the accuracy can certainly be
improved in forthcoming experiments.
\qpar

In addition to the improvements already suggested,
it is clear that the accuracy of the experiment can be
improved by increasing the number of trials. If instead of
5 trials, one gets results for 20 trials, the standard deviation of
the average will be divided by $ \sqrt{20/5}=2 $.
\qpar

Incidentally, Table 1 shows that the standard deviation
is almost always larger at noon than at midnight. For the
5 trials the average of the standard deviation is $ 16\mu s$ at noon
and $ 12\mu s $ at midnight.

\qI{Conclusions}

The three experiments considered 
in the present series of papers rely on two different
approaches. The approach of Parts 1 and 2 
consists in observing
the vectors of angular velocity associated with the movements of
rotation whereas in the approach of Part 3 one measures
gravitational forces. How do these approaches compare?
\qpar

The second approach depends upon several parameters: the mass and
distance of the body $ A $ which produces the attraction, 
the diameter of the body $ B $
(i.e. the Earth in our case) on which the measurement is performed%
\qfoot{Even its precise shape will play a role for instance
the fact that the Earth is not exactly a sphere; in contrast such
details are completely irrelevant in the angular velocity approach.}%
,
the distance between the centers of $ A $ and $ B $. 
As an illustration, let us for a moment assume that
the Earth has the size of Jupiter.
Its orbit around the Sun would be almost the same%
\qfoot{The angular velocity is given by Kepler's third law
which says that if the major axis of the orbit is the same then 
the orbital period will also be the same.}
but the tidal force due to the Sun would be much larger. 
On the contrary if the Earth were of the size of 
an asteroid such as Ceres (480 km) the tidal 
force would be very small. \qL
In addition, the tidal effect depends
upon internal properties of the Earth such as the rigidity or mobility
of its structural layers. This is the origin of the so-called gravific
factor, an empirical parameter of 
the order of 1.5, which has to be applied to the tidal potential
to get actual gravity changes.

\qpar

In contrast, the angular velocity approach does not depend 
on the diameter of $ B $
nor does it depend (directly) upon the distance $ AB $. 
The only parameter which really matters is the period of rotation.
Moreover the first approach allows the detection of rotations
(such as the rotation of the Earth around its axis or
its precessional axial rotation) 
which are not rotations
around another astronomical body%
\qfoot{Such movements can also be detected with a simple pendulum
although in such cases the detection relies 
on observing the centrifugal force
(rather than the tidal force) at two places located 
at different latitudes; see below for more details.}%
. 
In short, the first approach
seems to provide a broader and ``cleaner'' view.
\qpar

Of the three measurement methods that we have considered which one 
is the easiest to implement?
At the present moment it is still difficult to answer this question
because the most meaningful test, namely the detection of the 
rotation of the Earth around the Moon, has not yet been performed
by more than one method.
\qpar

The only comparison which can be made concerns the rotation of the
Earth around its axis. This rotation can of course be detected
with a Foucault pendulum. It can also be detected with a simple 
pendulum provided one compares measurements of the period performed
at different locations. Such an observation was made for the first
time in 1672 by the astronomer Jean Richer when he discovered
that a pendulum which has a period of 2\ s in Paris (at a latitude
of 49 degrees) has a period shorter by 3.4\ ms in the town of
Cayenne in French Guyana at a latitude of 4\ degrees%
\qfoot{The formula is the following:
$$ T\simeq T_0\left(1-{ \Delta g\over 2g }\right),\quad \hbox{where:}\quad
\Delta g=\Omega^2 R \cos\lambda,\quad \Omega={ 2\pi\over T },\quad
T=24\times 3600\ \hbox{s},\quad \lambda =\hbox{latitude} $$
}%
.
The fact that the Richer experiment took place almost two centuries
before the Foucault experiment suggests that somehow it was
easier to do. However, it required a long journey 
from Paris to the Equator.

\appendix

\qI{Appendix A: Formulas for tidal forces}

This appendix gives theoretical results for the magnitude
of the tidal effect. We are only concerned in the change of $ g $.
As the theory of tidal forces can
be found in many textbooks we focus mainly on a number of 
key-points.
\qpar

As in all problems of classical mechanics there are two
crucial preliminary steps (i) What simplifying assumptions do we make?
(ii) In which frame of reference do we work?

\qA{Simplifications}

We make the following simplifying hypotheses.
\qee{1} In a first step we concentrate on the Earth-Moon system and
forget the influence of the Sun. 
\qee{2} We assume that the orbit of the Moon is a circle so that the
Earth-Moon distance does not change in the course of time.
\qee{3} We assume that the Earth is a perfect sphere so that its
radius does not change with the latitude.

The most drastic simplification is the first one. However, once we have
been able to get the answer for the Earth-Moon system we will
observe that the argument is basically the same for the Earth-Sun
system.

\qA{Frame of reference}

The investigation of any phenomenon in classical mechanics 
requires that
one defines {\it two} frames of reference: one ($ F1 $ in which one will
work and one ($ F_0 $) that is considered as absolute%
\qfoot{An absolute (or inertial) frame of reference is one
in which one can apply Newton's law $ \qvec{F}=m\qvec{a} $
without introducing any additional corrective force such
as the centrifugal or Coriolis forces.}
and one ($ F_1 $) in which one will work. 
For many problems (for instance the acceleration of a car) the
frame of reference in which one works can be considered
as absolute even though it is not. In the example of the 
the car the corrective forces due to the fact that the
frame of reference rotates with the Earth are small with respect
to the forces which act on the car.
\qpar

For the tidal effect the frames $ F_0 $ and $ F_1 $ are given in Fig. A1.
Because $ F_1 $ rotates around the center of mass of the Earth-Moon
system, a correction must be introduced in the form of a centrifugal
force. Fig. A2 confirms that, as expected,
this force is the same for all points of the Earth%
\qfoot{For a pendulum the Coriolis force will be
perpendicular to the plane of oscillation of the pendulum.
As this pendulum 
has only one degree of freedom (instead of two for a Foucault
pendulum) the Coriolis force will play no role.}%
.

%
\begin{figure}[htb]
\centerline{\psfig{width=10cm,figure=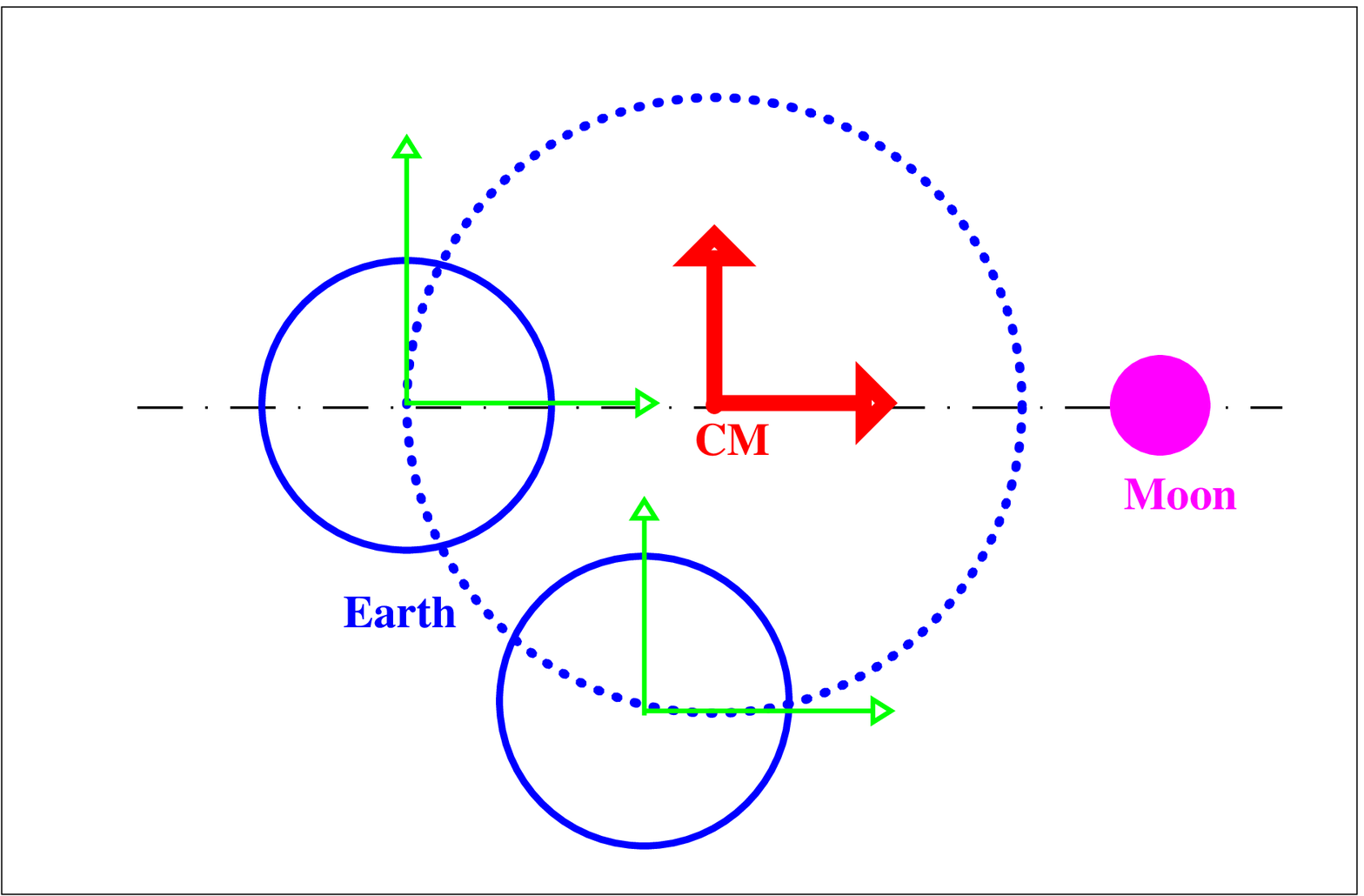}}
\vskip 3mm
\qleg{Fig.A1: Frames of reference for the Earth-Moon system.}
{For every problem in classical mechanics one must indicate
in which frame of reference one will work and which frame
of reference is considered as absolute 
that is to say steady except for a translation.\qL
The frame of reference in red ($ F_0 $) is considered as absolute. Its
origin is at center of mass of the 
Earth-Moon system and its axis have fixed directions with respect
to distant stars.\qL
The frame of reference in green ($ F_1 $)
is the one in which we will work.
Its origin is at the center of the Earth and the directions
of its axis are also fixed with respect to distant stars.
In this frame
the coordinates of the North Pole (for instance)
will remain unchanged while the Earth moves around the center of mass
which is what one wants for the sake of simplicity.\qL
As this frame of reference has a movement of rotation,
in order to be able to apply Newton's law one must
introduce two correcting forces: the centrifugal force and  the
Coriolis force. Because we will focus on the static aspect of the
tidal effect, the Coriolis force will play no role and we can forget it.}
{}
\end{figure}

%
\begin{figure}[htb]
 \centerline{\psfig{width=10cm,figure=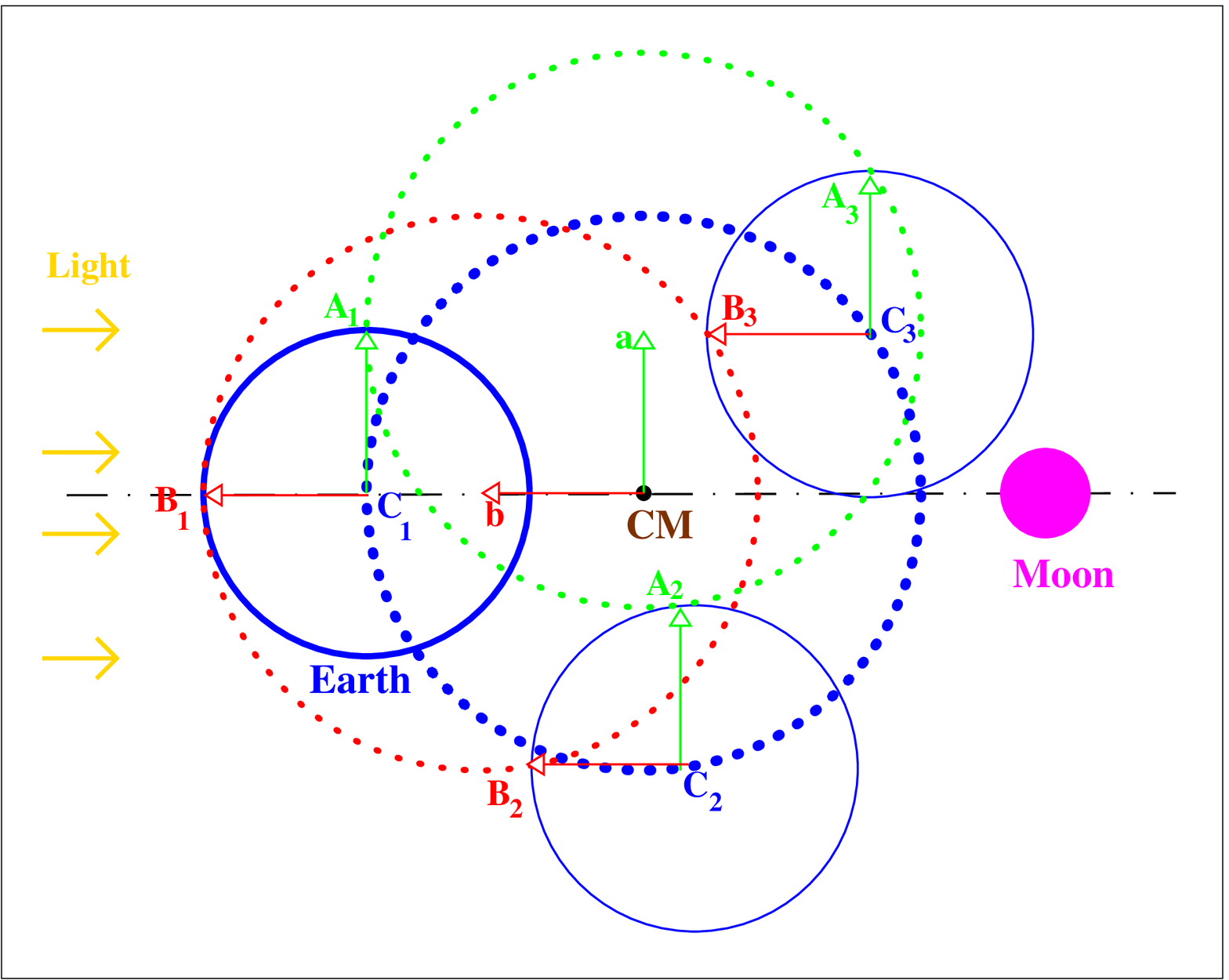}}
\vskip 3mm
\qleg{Fig.A2: Circular movement of the Earth around the center-of-mass
of the Earth-Moon system.}
{The purpose of the figure is to describe the movement of the Earth
around the center-of-mass of the Earth-Moon system when one assumes
that there is no rotation of the Earth itself around its center.
In this case all the points of the surface of the Earth move on 
circles which are identical to the trajectory of the center of the
Earth (blue dotted circle)
except for an appropriate translation. Thus, the North Pole
(points $ A_1, A_2, A_3 $) moves on the green circle centered on $ a $.\qL
For the sake of clarity of the figure the center of mass (CM)
has been represented between the Earth and the Moon whereas in fact it
is contained inside of the Earth at a distance of 4,600 km of its center.
}
{}
\end{figure}
%

\begin{figure}[htb]
\centerline{\psfig{width=10cm,figure=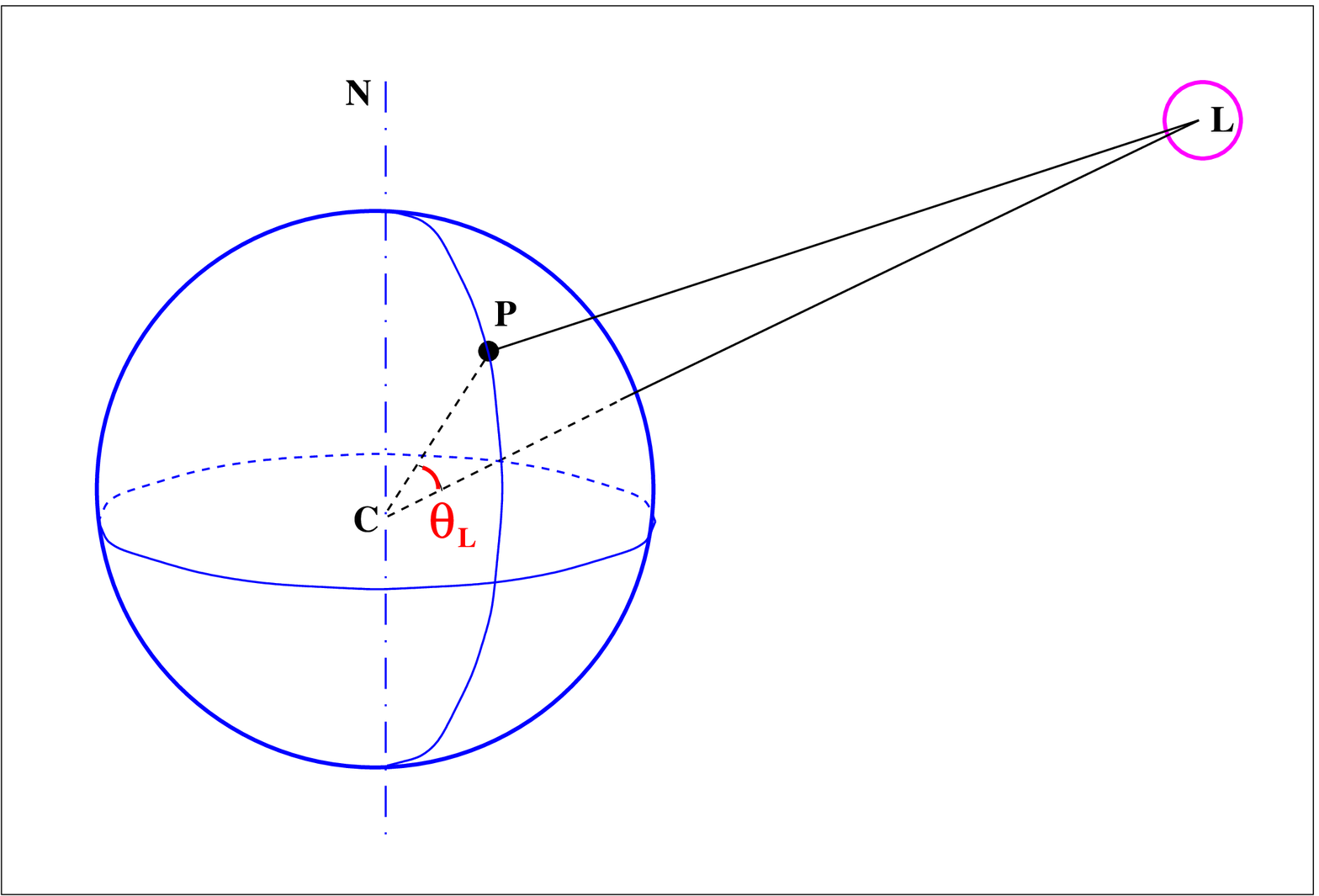}}
\vskip 3mm
\qleg{Fig.A3a: Definition of the angle $ \theta_L=\widehat{LCP} $.}
{The angle $ \theta_L $ between the vectors $ \qvec{CP} $ and
$ \qvec{CL} $ plays a key role in the formula of the tidal effect.
Fig. A3b shows how it can be expressed as a function of several
other angles which characterize the position of the observation
point $ P $ and the location of the Moon on its orbit around
the Earth.}
{}
\end{figure}
%
\begin{figure}[htb]
 \centerline{\psfig{width=10cm,figure=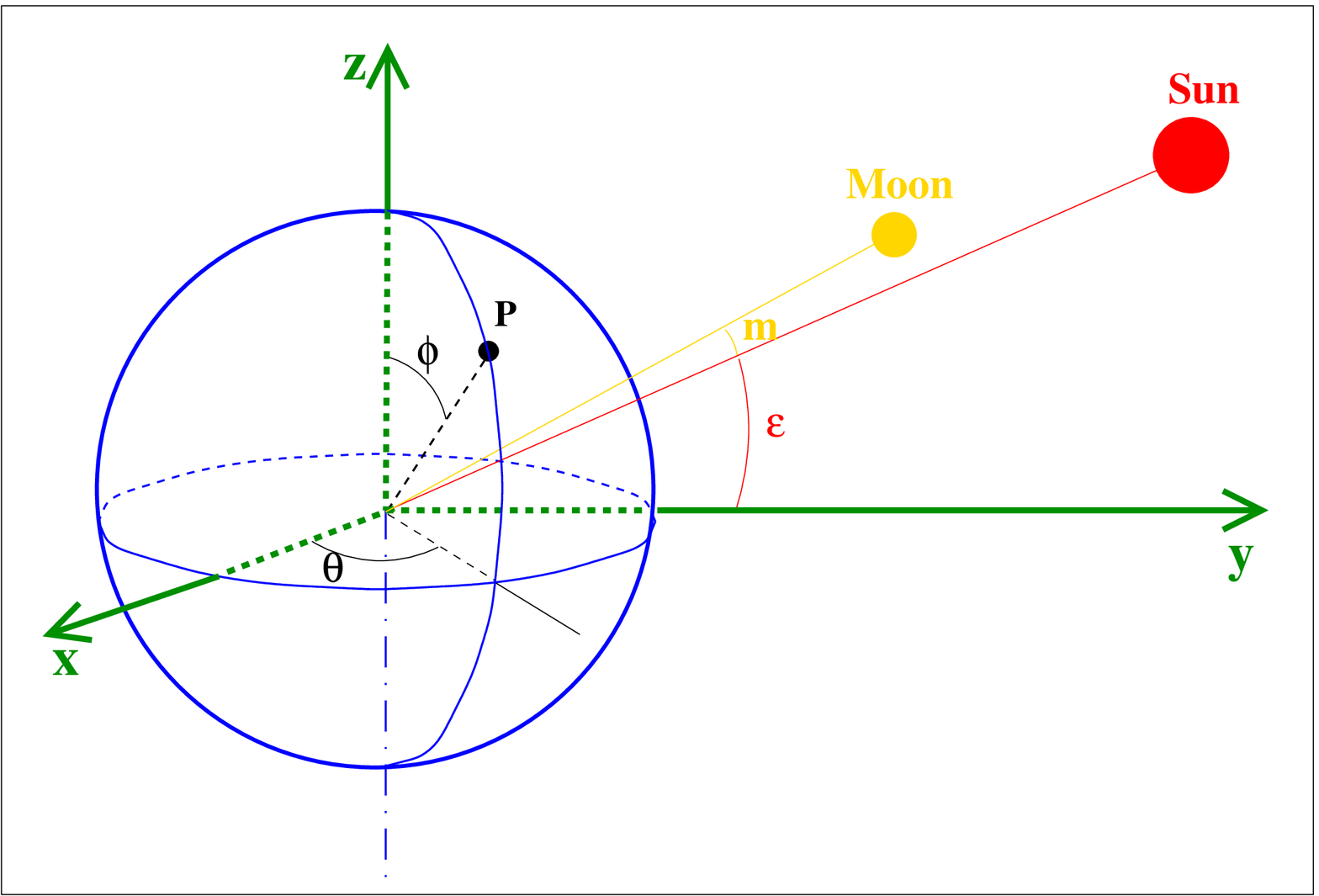}}
\vskip 3mm
\qleg{Fig.A3b: Definition of the frame of reference for
computing the angle $ \theta_L $.}
{The origin $ O $ of the frame of reference $ Oxyz $ is at the center
of the Earth, $ Oz $ is on its axis of rotation, $ Oy $ is in the
direction of the Sun. The location $ P $ is defined by the 
two spherical angles $ \theta, \phi $. The figure shows the
Moon in ``new moon'' position; when the Moon moves on its orbit
it will be described by an angle $ \mu $ with respect to $ Ox $
in the plane which contains $ Ox $ and the Moon.}
{}
\end{figure}

%
\begin{figure}[htb]
 \centerline{\psfig{width=10cm,figure=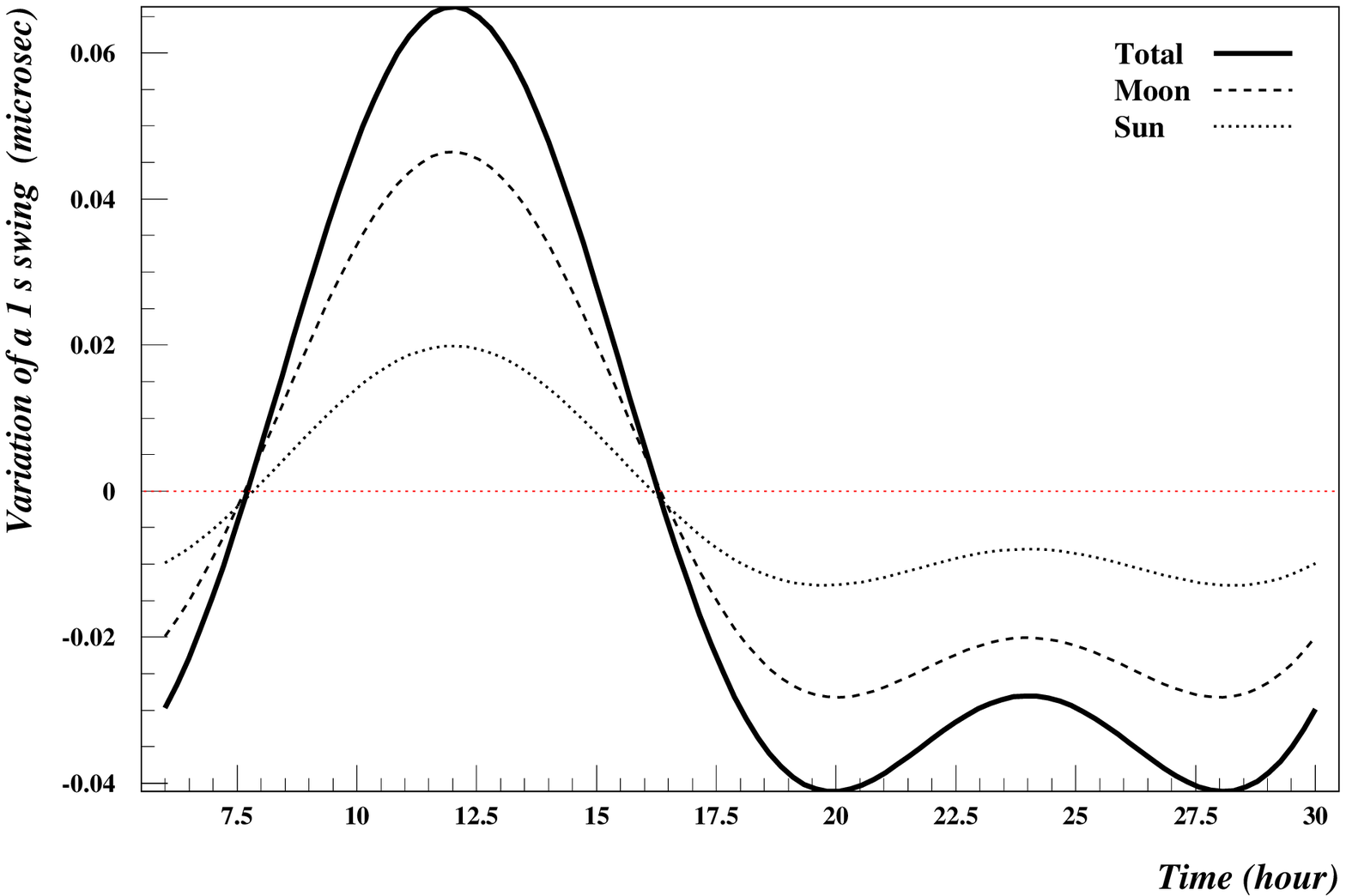}}
\vskip 3mm
\qleg{Fig. A4: Theoretical variations of the beat of a pendulum
(new moon, April).}
{The horizontal scale gives the time starting from 6 am to 6 am
on the next day which on the graph corresponds to time
6+24=30:00. The latitude of the place is 46 degree.
In this configuration
the difference between noon and midnight is $ \Delta T'=0.094\mu $s;
incidentally, 
the difference between noon and 8 pm is slightly larger, namely.
$ 0.106\mu $s. The shape of the curve
would be almost the same in an
other month (provided that one is a new moon configuration)
but the horizontal time scale would be different.
In June for instance one gets $ \Delta T'=0.098\mu $s.
On the contrary, in a first (or last) quarter configuration
the difference is much smaller; thus, for a (first quarter, June)
configuration one gets: $ \Delta T'=0.028\mu $s.
Some rules providing quick estimates of the tidal force
are given in the text.}
{}
\end{figure}

\qA{Formulas}

The three formulas which are needed for computing the tidal
effect on the beat of a pendulum are the following%
\qfoot{Formula (A1) comes from Leroy (2004, p. 12).}%
:
$$ p=a_v/g= { M_L\over M_T }\left({ R\over d }\right)^3
[3\cos^2\theta_L -1] \qn{A1} $$

where:\qL
$ a_v $: Vertical component of the tidal force\qL
$ g $: Acceleration of gravity \qL
$ M_L $: Mass of the Moon \qL
$ M_T $: Mass of the Earth \qL
$ R $: Radius of the Earth \qL
$ d $: Distance from the Earth to the Moon \qL
$ \theta_L $: Angle defined in Fig. A3a
\qpar

The angle $ \theta_L $ is defined by:
$$ \cos\theta_L=\cos\mu \sin\phi \cos\theta + 
 \sin\mu \cos\alpha \sin\phi \sin\theta + 
 \sin\mu \sin\alpha \cos\phi \qn{A.2} $$

where:\qL
$ \theta_L $: Angle defined in Fig. A3a \qL
$ \mu $: Angle between $ Ox $ and the Moon in the plane which
 contains both $ Ox $ and the Moon \qL
$ \theta,\ \phi $: Spherical angles as defined in Fig. A3b;
$ \phi $ is the co-latitude $ \pi/2-\lambda $ where 
$ \lambda $ is the latitude of $ P $; $ \theta $ increases
as the Earth turns around its axis. With the frame of reference
of Fig. A3b, $ \theta=\pi/2 $ at noon and 
consequently $ \theta=0 $ at 6 am \qL
$ \alpha $: Angle between the normal to the plane which contains
the orbit of the Moon and $ Oz $; on 21 June (situation
shown in Fig. A3b) $ \alpha=\epsilon +m = 23+5=28 $ degrees.
On 21 December, $ \alpha=\epsilon -m = 18 $ degrees. On
other dates $ \alpha $ is comprised between these limits
and given by the formula: 
$$ \cos\alpha =[\sin e \sin\epsilon +
(1/m)\cos\epsilon](1+1/m^2)^{-1/2} $$
where $ e $ is the angle that determines the position of the Earth
on its orbit with respect to its position on 21 March.
\qpar

Finally, the change in the duration of a
beat $ T'=T/2 $ of a pendulum is given by:
$$ T'=\pi\sqrt{ { L\over g' } } =\pi\sqrt{ { L\over g(1+p) } }
=T'_0(1-p/2) \qn{A.3} $$

The graph in Fig. A.4 illustrates the use of the above
formulas in a specific case.

\qA{Qualitative rules}

In order to get quick estimates for the tidal
effect at a location $ P $ one can use the following
rules. 
\qee{1}
As the tidal effect is nonexistent at the center ($ C $) of the 
Earth, it is the distance $ CP' $ which matters, where $ P' $
is the projection of $ P $ on the ecliptic.
\qee{2}
For places whose ``latitude'' $ \lambda_e $
with respect to the ecliptic
(instead of equator) is near 90 degree the tidal force is mainly
downward that is to say $ \Delta g > 0 $; for places for which
$ \lambda_e \sim 60 $ degree, the tidal force is almost horizontal.
which makes its vertical projection (the only one
which matters for a pendulum) fairly small.
\qee{3}
The difference in the period for
noon and midnight is largest for new moon
or full moon. 
\qee{4}
For almost all latitudes the effect of the Sun is about one
half the effect of the Moon. However, this is not true for
places where the two forces are fairly horizontal because in such cases
even a small difference in direction
can produce a substantial change for the vertical component.
\vskip 10mm

{\bf \large References}
\vskip 5mm

\qparr
B\'etrisey (M.) 2011: Radiometric clocks. \qL
http://www.betrisey.ch/eindex.htm. 

\qparr
Cabannes (H.) 1962, 1966: Cours de m\'ecanique g\'en\'erale.
Dunod, Paris.

\qparr
Goldstein (H.), Poole (C.), Safko (J.) 2002: Classical
mechanics. 3rd edition. Addison-Wesley.

\qparr
Leroy(J.) 2004: Notes sur l'origine physique de la mar\'ee.\qL
http://www.jleroy.net/Site/MAREES\_2.4.pdf

\qparr
Michelson (A.A.) 1914: Preliminary results for the measure of 
the rigidity of the Earth.
Astrophysical Journal, Vol. 39, 105-128.

\end{document}